\documentclass[a4paper,11pt]{article}
\usepackage{jheppub} 

\usepackage{graphicx} 
\usepackage{multirow}%
\usepackage{amsmath,amssymb,amsfonts}%
\usepackage{amsthm}%
\usepackage{mathrsfs}%
\usepackage[title]{appendix}%
\usepackage{xcolor}%
\usepackage{textcomp}%
\usepackage{manyfoot}%
\usepackage{booktabs}%
\usepackage{algorithm}%
\usepackage{algorithmicx}%
\usepackage{algpseudocode}%
\usepackage{listings}\usepackage{rotating} 
\usepackage[normalem]{ulem}
\usepackage{caption}

\arxivnumber{2407.10948} 

\title{\boldmath The infamous 95\,GeV b\=b excess at LEP: \\ Two b or not two b?}

\author{Patrick Janot\footnote{LEP Coordinator in 1999 and 2000}}
\affiliation{CERN, EP Department\\
1 Esplanade des Particules, 1217 Meyrin, Switzerland}

\emailAdd{Patrick.Janot@cern.ch}

\abstract{A small deviation observed around 95\,GeV in the diphoton invariant mass distribution in the LHC Run\,2 data has been subject to considerable attention in the past couple of years. The interpretation of this excess as the manifestation of an additional scalar particle at this mass is often claimed to be supported by a previously observed, even smaller, excess in the b\=b invariant mass distribution in LEP data. This short note aims at confronting this claim to factual experimental observations, through a careful scrutiny of the detailed LEP public notes written at the time on the topic. It is found that the LEP data strongly disfavour the production of a new 95\,GeV scalar particle, as well as any other new physics interpretation in the 95-100\,GeV mass range.}

\begin{document}
\maketitle
\flushbottom

\section{Introduction}
\label{sec:Intro}

The CMS and ATLAS collaborations recently released the result of their search for diphoton resonances with masses below 110\,GeV with the full Run\,2 data sample~\cite{CMS:2024yhz,ATLAS:2024bjr}. One of the motivations for this search was a $2.9\sigma$ excess observed in the CMS diphoton mass distribution at 95.3\,GeV in 2016 with approximately a quarter of the Run\,2 data~\cite{CMS:2018cyk}. 
The significance of the excess did not increase with the four times larger CMS data sample recorded from 2016 to 2018, nor did it increase with an eight-fold increase of the statistics achieved in an unofficial combination of the CMS and the ATLAS Run\,2 data~\cite{Biekotter:2023oen}. Should the 2016 CMS excess originate from an actual new physics signal, a more impressive $2.9\sigma \times \sqrt{8} = 8.2\sigma$ significance would have naively been expected by now at Run\,2. 

This ill-behaved $\sim 3\sigma$ (local) significance has, however, triggered a large number of phenomenological preprints and publications~\cite{Biekotter:2022jyr,Biekotter:2023jld,Azevedo:2023zkg,Escribano:2023hxj,
Belyaev:2023xnv,Ashanujjaman:2023etj,Aguilar-Saavedra:2023tql,Dutta:2023cig,Ellwanger:2023zjc,Cao:2023gkc,Borah:2023hqw,Ahriche:2023hho,Arcadi:2023smv,Ahriche:2023wkj,Chen:2023bqr,Dev:2023kzu,Li:2023kbf,Belyaev:2024lah,Liu:2024cbr,Wang:2024bkg,Cao:2024axg,Kalinowski:2024uxe,Ellwanger:2024txc,Kalinowski:2024oyy,Diaz:2024yfu,Ellwanger:2024vvs,Ayazi:2024fmn,Arhrib:2024wjj,Benbrik:2024ptw,Ge:2024rdr,Yang:2024fol,Lian:2024smg,Mosala:2024mcy,Gao:2024qag} in the past two years, in an attempt of  exhibiting more or less credible new physics explanations for this 95\,GeV excess.\footnote{Earlier similar papers~\cite{Biekotter:2019kde,Kundu:2019nqo,Cao:2019ofo,Aguilar-Saavedra:2020wrj,Biekotter:2020cjs,Heinemeyer:2021msz,Benbrik:2022azi} alluded to a 96\,GeV scalar instead, following the movements of the CMS excess.} To reinforce the limited strength of the observed ``signal'' and psychologically increase the substance of their claims, all these papers refer to a $2.3\sigma$ deviation seen around 98\,GeV in the SM Higgs boson search at LEP at the end of the previous millennium~\cite{LEPWorkingGroupforHiggsbosonsearches:2003ing}. This small excess seen in the $\rm b\bar b$ (but not in the $\tau^+\tau^-$) invariant mass distribution is consistently used in support of the $3\sigma$ deviation in the LHC diphoton mass distribution at 95\,GeV. 

The argumentation used in these papers to justify this support is based on the claims from Refs.~\cite{Biekotter:2022jyr,Biekotter:2023jld}, the language of which has been  systematically propagated from one paper to the next without questioning the robustness of the reasoning. In a first leap of faith, it is stated that the small excess of $\rm b\bar b$ events at LEP hints at the production of a new scalar particle $\phi$ with mass of approximately 98\,GeV through the process $\rm e^+e^- \to Z\phi \to Z b\bar b$, with a signal strength (i.e., the ratio to the corresponding SM Higgs cross section) of $0.117 \pm 0.057$, as inferred from Ref.~\cite{Cao:2016uwt}. In a second step, it is argued that the dijet mass resolution at LEP would be so much worse than the diphoton mass resolution at LHC that the 98\,GeV LEP excess could actually correspond to the putative 95\,GeV particle responsible for the diphoton excess at LHC.\footnote{About half of the papers~\cite{Aguilar-Saavedra:2023tql,Ahriche:2023wkj,Ahriche:2023hho,Arcadi:2023smv,Arhrib:2024wjj,Ashanujjaman:2023etj,Ayazi:2024fmn,Azevedo:2023zkg,Belyaev:2023xnv,Belyaev:2024lah,Benbrik:2024ptw,Biekotter:2022jyr,Biekotter:2023jld,Biekotter:2023oen,Borah:2023hqw,Cao:2023gkc,Cao:2024axg,Chen:2023bqr,Dev:2023kzu,Diaz:2024yfu,Dutta:2023cig,Ellwanger:2023zjc,Ellwanger:2024txc,Ellwanger:2024vvs,Escribano:2023hxj,Ge:2024rdr,Kalinowski:2024oyy,Kalinowski:2024uxe,Li:2023kbf,Lian:2024smg,Liu:2024cbr,Wang:2024bkg,Yang:2024fol,Mosala:2024mcy,Gao:2024qag} do not mention the 98\,GeV value and directly allude to a LEP $\rm b\bar b$ excess at 95\,GeV, or remain vague, referring to the ``95--100\,GeV region''.}

It is the purpose of this short note to expose the flaws of this reasoning. This note is organised as follows. In Section~\ref{sec:mass}, the dijet mass determination in the LEP Higgs boson search is recalled, and it is shown that its resolution is not that different from the diphoton mass resolution at LHC.  A brief description of the LEP data collected between 1998 and 2000, relevant for the search of a 95\,GeV Higgs boson, is given in Section~\ref{sec:data}. The LEP conference notes and publications are scrutinised in Sections~\ref{sec:189} and~\ref{sec:192-210}. This scrutiny leads to the conclusion that the LEP data cannot be used in support of the production of a scalar particle with mass 95\,GeV, as these data {\it (i)} do not show any consistent evidence for such a particle, and {\it (ii)} actually strongly disfavour this hypothesis.  A summary of these findings is given in Section~\ref{sec:conclusion}. 

\section{Dijet mass resolution in the LEP Higgs boson search}
\label{sec:mass}

The simplest way to measure a dijet mass (called ``raw'' dijet mass in the following) is to use the measured energies and momenta of the particles reconstructed and identified in the LEP detectors, and belonging to one of the two jets. For example, with the sophisticated particle-flow reconstruction in the ALEPH detector~\cite{ALEPH:1994ayc}, the resolution on the raw dijet mass $M$ (averaged over all quark flavours) is parameterised as follows:
\begin{equation}
\sigma(M) = \left( 0.59 \pm 0.03\right) \sqrt{M/{\rm GeV}} + \left( 0.6 \pm 0.3\right) {\rm GeV},
\end{equation}
\noindent corresponding to a relative resolution of about 7\% for a dijet mass of 95\,GeV: this is indeed significantly worse than the 1 to 2.5\% resolution of the diphoton mass in the CMS detector~\cite{CMS:2024yhz}. The raw $\rm b \bar b$ mass measurement is affected in addition by undetected neutrinos from frequent semi-leptonic decays, which further degrade the resolution to about 10\%. It would a priori require at least 100 events to reach a 3$\sigma$ separation between a 98\,GeV and a 95\,GeV particle. It is assumed here that the claim of Ref.~\cite{Biekotter:2022jyr,Biekotter:2023jld} regarding the ``coarse/limited $\rm b\bar b$ mass resolution at LEP'' is based on this reasoning.  

The search for the Higgs boson at LEP, however, proceeds by the analysis of the $\rm e^+e^- \to ZH \to Z b \bar b$ process, the kinematics of which is highly constrained for the four total energy and momentum conservation equations and, for the non-leptonic Z decays, by the knowledge of the Z mass. These constraints decisively improves the $\rm b \bar b$ mass resolution with respect to the raw mass, in a way that depends on the Z decay channel~\cite{Boucrot:1987qv}, as described below. 
\begin{enumerate}
\item 
When the Z decays into a pair of charged leptons $\ell^+\ell^-$ ($\ell = \rm e , \mu$), the $\rm b \bar b$ mass is defined as the mass recoiling against the very-well measured lepton pair in the context of total energy/momentum conservation, without even bothering with the Higgs boson decay products: 
\begin{equation}
m^2_{\rm b \bar b} = s + m^2_{\ell\ell} - 2\sqrt{s}(E_{\ell^+} + E_{\ell^-}),
\end{equation}
\noindent where $m_{\ell\ell}$ is the measured lepton pair invariant mass, $E_{\ell^+}$, $E_{\ell^-}$ are the two measured lepton energies, and $\sqrt{s}$ is the centre-of-mass energy. In the ALEPH detector, the Higgs boson mass relative resolution achieved in this channel is of the order of 1\%. 
\item 
When the Z decays instead into a pair of quarks $\rm q \bar q$ or a pair of taus $\tau^+\tau^-$, leading to a four-jet final state, the energies of the Z decay products are not measured well enough and the recoil method is thus no longer very precise. The four jet energies, however, can be recomputed from the very accurately measured jet directions (or velocities) $\vec{\beta}_i = \vec{p}_i^{\rm \, raw}/E_i^{\rm raw}$, where $\vec{p}^{\rm \,raw}_i$, $E_i^{\rm raw}$ are the measured momentum and energy of jet $i$, by solving for $E_1, E_2, E_3, E_4$ the four (linear) total energy-momentum conservation equations: 
\begin{eqnarray}
\vec{\beta}_1 E_1 + \vec{\beta}_2 E_2 + \vec{\beta}_3 E_3 + \vec{\beta}_4 E_4 & = & \vec{0}, \\
\phantom{\vec{\beta}_1} E_1 + \phantom{\vec{\beta}_2} E_2 + \phantom{\vec{\beta}_3} E_3 + \phantom{\vec{\beta}_4} E_4 & = & \sqrt{s},
\end{eqnarray}
allowing the dijet invariant masses $m_{12}$ (for the Z) and $m_{34}$ (for the Higgs boson) to be more precisely determined from the recomputed jet energies $E_1$, $E_2$, $E_3$ and $E_4$, obtained by a straightforward $4\times 4$ matrix inversion. In this four-jet final state, the Z mass constraint is applied as follows
\begin{equation}
m_{\rm b \bar b} = m_{12} + m_{34} - m_{\rm Z},
\end{equation}
and yields an even more accurate reconstructed $\rm b \bar b$ mass determination  with a typical relative core resolution of the order of 2\% in the ALEPH detector. 
\item
Finally, when the Z decays into a pair of neutrinos $\nu\bar\nu$, the only remaining constraint is that the missing mass should equal $m_{\rm Z}$. Applying this constraint yields a second-order polynomial equation, which solves as: 
\begin{equation}
m_{\rm b\bar b} = \frac{\sqrt{s}E_{\rm raw}-\sqrt{sE^2_{\rm raw}-m^2_{\rm raw}\left(s-m^2_{\rm Z}\right)}}{m_{\rm raw}}, 
\end{equation}
where $E_{\rm raw}$ and $m_{\rm raw}$ are the raw total energy and mass of the visible final state particles measured in the detector. In the ALEPH detector, the typical reconstructed $m_{\rm b\bar b}$ relative resolution achieved in that case is of the order of 3\%. \end{enumerate}

In summary, the total energy, total momentum and Z mass constraints -- when applied to the ZH final state kinematics -- improves the $\rm b \bar b$ mass relative core resolution from 10\% to 1--3\% in the ALEPH detector.\footnote{When averaged over the four LEP detectors and all channels, the $\rm b\bar b$ mass relative core resolution is slightly worse than that of the sole ALEPH detector. This remark does not affect the conclusions of this note. A core resolution of $\sigma_{\rm core}^{\rm LEP} = 3$\,GeV is conservatively assumed throughout.} This kinematically constrained $\rm b\bar b$ mass core resolution is remarkably similar to the 1--2.5\% diphoton raw mass resolution in the CMS detector (for which no kinematic constraint exists). With this resolution, a handful of events in the core of the signal mass distribution over a relatively small background would suffice to unambiguously distinguish the $\rm b\bar b$ invariant mass distributions resulting from the decay of either a 95\,GeV or a 98\,GeV particle, in contrast with the hypothesis made implicitly or explicitly in Refs.~\cite{Biekotter:2023oen,Biekotter:2022jyr,Biekotter:2023jld,Azevedo:2023zkg,Escribano:2023hxj,
Belyaev:2023xnv,Ashanujjaman:2023etj,Aguilar-Saavedra:2023tql,Dutta:2023cig,Ellwanger:2023zjc,Cao:2023gkc,Borah:2023hqw,Ahriche:2023hho,Arcadi:2023smv,Ahriche:2023wkj,Chen:2023bqr,Dev:2023kzu,Li:2023kbf,Belyaev:2024lah,Liu:2024cbr,Wang:2024bkg,Cao:2024axg,Kalinowski:2024uxe,Ellwanger:2024txc,Kalinowski:2024oyy,Diaz:2024yfu,Ellwanger:2024vvs,Ayazi:2024fmn,Arhrib:2024wjj,Benbrik:2024ptw,Ge:2024rdr,Yang:2024fol,Lian:2024smg,Mosala:2024mcy,Gao:2024qag}.

To understand the next sections, it is important to note that, beside the core mass resolution, non-Gaussian tails may occur for several reasons, such as inaccuracies in the jet clustering, incorrect jet assignment, or hard initial/final state radiation, making the reconstructed final state less compatible with the ZH kinematics. Also, the Z mass constraint imposes that the inferred $\rm b\bar b$ mass is always smaller than the ZH production threshold $\sqrt{s}-m_{\rm Z}$. The $\rm b\bar b$ mass distribution predicted for a particle with mass close to (or larger than) this threshold will therefore be skewed towards smaller values. As a consequence, and while the narrow core of the mass distribution would allow the mass of a hypothetical new particle to be determined with a GeV precision from a handful of events, each event, when taken individually, has a nonzero probability to have a true mass far from the kinematically constrained $m_{\rm b \bar b}$ value. An excess at any reconstructed mass therefore affects the signal-to-noise likelihood ratios and the confidence levels over broad ranges of particle mass hypotheses, whether it comes from a true signal or from a background fluctuation. A true signal, however, comes with expected characteristic likelihood ratio behaviour, invariant mass distribution, and centre-of-mass energy dependence, which need to be confronted with the data to ascertain the signal interpretation. This basic sanity check is definitely missing in Ref.~\cite{Biekotter:2022jyr,Biekotter:2023jld} and in the following incarnations~\cite{Biekotter:2023oen,Azevedo:2023zkg,Escribano:2023hxj,
Belyaev:2023xnv,Ashanujjaman:2023etj,Aguilar-Saavedra:2023tql,Dutta:2023cig,Ellwanger:2023zjc,Cao:2023gkc,Borah:2023hqw,Ahriche:2023hho,Arcadi:2023smv,Ahriche:2023wkj,Chen:2023bqr,Dev:2023kzu,Li:2023kbf,Belyaev:2024lah,Liu:2024cbr,Wang:2024bkg,Cao:2024axg,Kalinowski:2024uxe,Ellwanger:2024txc,Kalinowski:2024oyy,Diaz:2024yfu,Ellwanger:2024vvs,Ayazi:2024fmn,Arhrib:2024wjj,Benbrik:2024ptw,Ge:2024rdr,Yang:2024fol,Lian:2024smg,Mosala:2024mcy,Gao:2024qag}, but is carefully addressed in the following sections.      

\section{The LEP data relevant to a \texorpdfstring{95\,GeV}{95} scalar production}
\label{sec:data}

In $\rm e^+e^-$ collisions, the production of a 95\,GeV scalar particle in association with a Z boson becomes significant for centre-of-mass energies $\sqrt{s}$ in excess of the kinematic threshold $95+91=186$\,GeV. As indicated in Table~\ref{tab:LEPLumis}, LEP overtook this threshold in 1998, with a regular $\sqrt{s}$ increase until the end of 2000 all the way to 209.2\,GeV. Table~\ref{tab:LEPLumis} gives the total integrated luminosities collected at LEP at each of the (averaged) centre-of-mass energies above this threshold, together with the cross section of the $\rm e^+e^- \to Z\phi \to Z b\bar b$ process for  $m_{\rm \phi} = 95$\,GeV for a signal strength of 0.117; the cross section of the dominant $\rm e^+e^- \to ZZ \to Z b \bar b$ background; and, for illustration, the SM ZH cross section with $\rm H \to b\bar b$ for  $m_{\rm H} = 98$\,GeV. The corresponding numbers of signal and background events expected to be produced in 1998, 1999, and 2000 are given in the last three rows.

\begin{table}[htbp]
    \renewcommand{\arraystretch}{1.3}
    \centering
    \caption{\small Integrated luminosities collected by the four LEP experiments in the last three years of operation, at centre-of-mass energies of 189\,GeV and above. The cross sections of the $\rm e^+e^- \to Z\phi \to Z b\bar b$ process for $m_{\rm \phi} = 95$\,GeV for a signal strength of 0.117, calculated with the HZHA program~\cite{hzha}; the dominant $\rm e^+e^- \to ZZ \to Z b \bar b$ background~\cite{Mele:456346}; and, for illustration, of the SM Higgs production with $\rm H \to b \bar b$ for $m_{\rm H} = 98$\,GeV;  as well as the corresponding numbers of signal and background events expected to be produced in 1998, 1999, and 2000 are given in the last six rows.}
    \begin{tabular}{|c|c|c|c|c|c|c|c|}
    \hline
    Year & 1998 & \multicolumn{4}{|c|}{1999} & \multicolumn{2}{|c|}{2000} \\ \hline
    $\sqrt{s}$ (GeV) & 189 & 192 & 196 & 200 & 202 & 205 & $\ge 206$ \\ \hline
    L (pb$^{-1}$) & 683 & 113 & 315 & 331 & 156 & 325 & 536 \\ \hline 
    $\sigma^{0.117}_{95}$ (pb) & 0.017 & 0.025 & 0.031 & 0.034 & 0.036 & 0.037 & 0.038 \\ \hline
    $\sigma_{\rm Z b\bar b}$ (pb) & 0.182 & 0.215 & 0.252 & 0.274 & 0.282 & 0.291 & 0.296 \\ \hline
    $\sigma^{\rm SM}_{98}$ (pb) & 0.049 & 0.131 & 0.203 & 0.246 & 0.261 & 0.278 & 0.284 \\ \hline\hline
    $N^{0.117}_{95}$ & 12 & \multicolumn{4}{|c|}{29} & \multicolumn{2}{|c|}{32} 
    \\ \hline
    $N_{\rm Z b\bar b}$ & 124 & \multicolumn{4}{|c|}{238} & \multicolumn{2}{|c|}{253} 
    \\ \hline
    $N^{\rm SM}_{98}$ & 34 & \multicolumn{4}{|c|}{201} & \multicolumn{2}{|c|}{242} 
    \\ \hline
    \end{tabular}
    \label{tab:LEPLumis}
\end{table}


Conference notes giving relevant details about the SM Higgs boson search at LEP were submitted in 1999~\cite{Bock:411797} for the 1998 data at $\sqrt{s} = 189$\,GeV; in 2000~\cite{Carena:436918} for the 1999 data at $192 \le \sqrt{s} \le 202$\,GeV; and in 2001~\cite{alephcollaboration2001search}, for the 2000 data at $200 \le \sqrt{s} \le 210$\,GeV, respectively. The year 2000 started with collisions at $\sqrt{s} = 200$ and 202\,GeV for a week or so, but the vast majority of the 2000 data were collected above 205\,GeV, culminating at 209.2\,GeV in the very last days. These data are labelled ``200--210\,GeV'' in some of the figures of this note. The final overall combination is presented in Ref.~\cite{LEPWorkingGroupforHiggsbosonsearches:2003ing}.

\section{The LEP excess at \texorpdfstring{$\sqrt{s} = 189$\,GeV}{189}}
\label{sec:189}

\begin{figure}[htbp]
    \centering
    \includegraphics[width=0.60\textwidth]{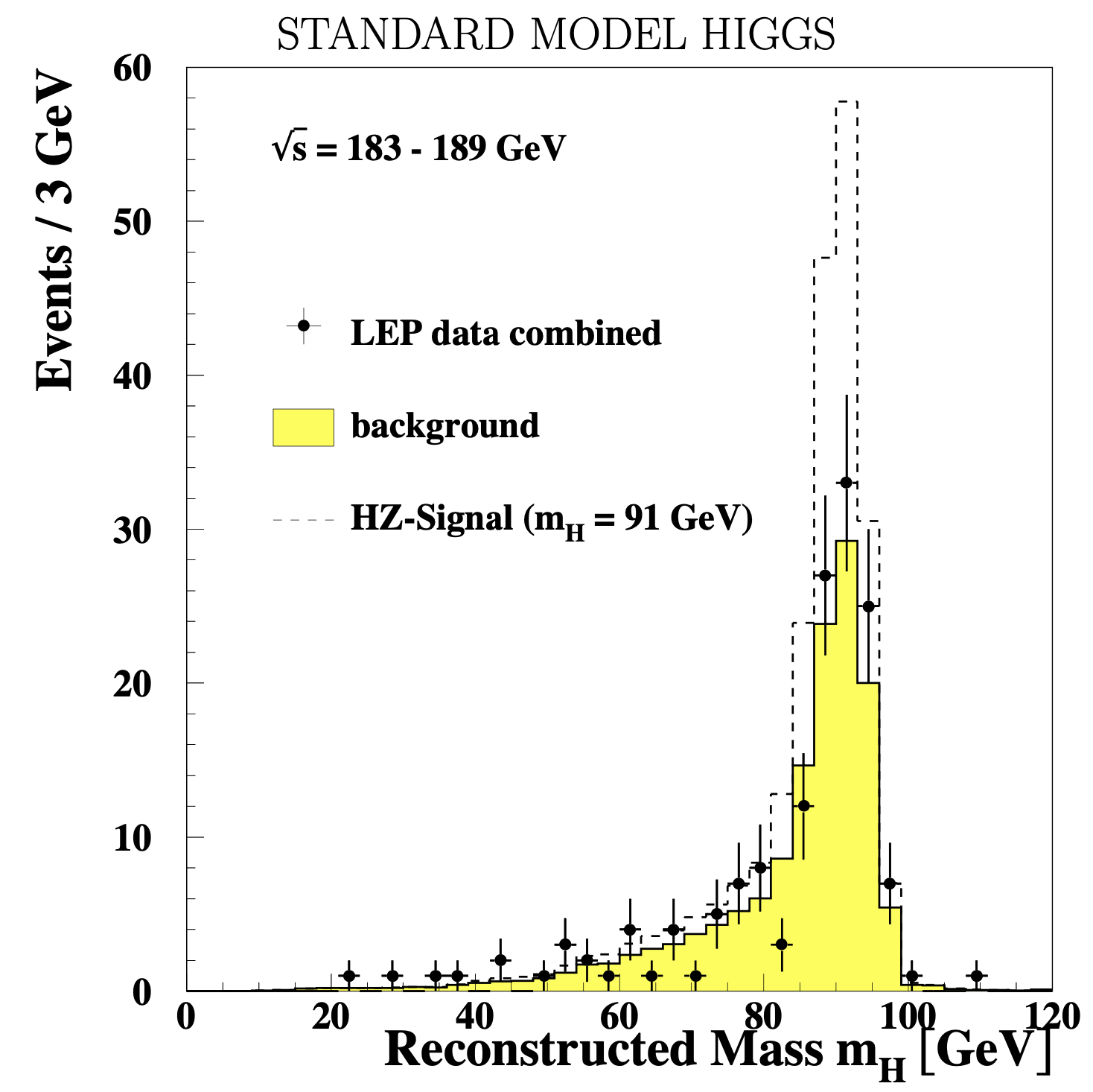}
    \caption{\small From Ref.~\cite{Bock:411797}: Distribution of the {\it reconstructed} Higgs boson mass in searches from the SM Higgs boson. The figure displays the data (dots with error bars), the predicted SM background (shaded histogram) and the prediction for a SM Higgs boson of 91\,GeV mass (dashed histogram). The distributions combine the data of all four LEP experiments at 189\,GeV plus the OPAL data at 183\,GeV. The number of data events entering this figure is 153 for 139.8 expected. A SM signal at 91\,GeV would amount to 89 events. The events entering these distributions were selected as follows. After a loose pre-selection in each experiment, the events were ranked according to their signal purity, determined from a discriminating variable that combines many features of the events (the most important of which being variables which tag b-flavoured jets) to distinguish on a statistical basis events from background processes and events from a Higgs boson signal. An experiment-dependent lower cut was then applied on this signal purity, thus rejecting the most background-like events, in order to get an equal signal-to-background ratio in all four experiments.  }
    \label{fig:ObsMass189}
\end{figure}

The raison-d'\^etre of this note is almost entirely based on a small excess of events in the 1998 LEP data. The distribution of the {\it reconstructed} mass in searches for the SM Higgs boson at $\sqrt{s} = 189$\,GeV from the four LEP experiments is displayed in Fig.~\ref{fig:ObsMass189}, taken from Ref.~\cite{Bock:411797}. A small overall excess of 13 events with respect to the SM background is indeed observed (153 events observed for 139.8 expected). This excess is mostly concentrated at the highest masses, with 66 events observed for 54 expected (mostly from the ZZ background) in a 12\,GeV window above 90\,GeV, corresponding to an excess of 12 events. The probability for this excess to arise from a background fluctuation (i.e, in the absence of signal) is estimated to amount to about 1\% in Ref.~\cite{Bock:411797}, for {\it test} masses from 94 to 98\,GeV, corresponding locally to 2.3 standard deviations. 

The predictions for a 95\,GeV or a 98\,GeV Higgs boson are not shown in Fig.~\ref{fig:ObsMass189}, but the prediction for a 91\,GeV SM Higgs boson is displayed above the dominant $\rm e^+e^- \to ZZ$ background (confirming the excellent reconstructed mass resolution at LEP). A signal at this mass would amount to 89 selected events out of 174 events expected to be produced with $\rm H\to b\bar b$. Of these 89 events, 74 would be expected to have a reconstructed mass in a 12\,GeV interval (i.e., $\pm 2\sigma^{\rm LEP}_{\rm core}$) around 91\,GeV, corresponding to a selection efficiency of approximately 42\% in this mass range. Assuming a mild dependence of the selection efficiency with the Higgs boson mass, and from the numbers of produced events displayed in Table~\ref{tab:LEPLumis}, about 5 events would thus be selected in a 12\,GeV window above 90\,GeV for a 95\,GeV signal with strength 0.117, while 14 events would be selected for a SM Higgs boson of mass 98\,GeV in the same interval. These numbers are to be compared with the observed excess of 12 events with a reconstructed mass above 90\,GeV. Given the expected background rate, and the fact that 98\,GeV is exactly at the kinematic threshold for ZH production at $\sqrt{s} = 189$\,GeV, thus biasing the signal reconstructed mass distribution to lower values, these 12 events would not suffice to distinguish it from a 95\,GeV signal.  

\begin{figure}[htbp]
    \centering
    \includegraphics[width=0.80\textwidth]{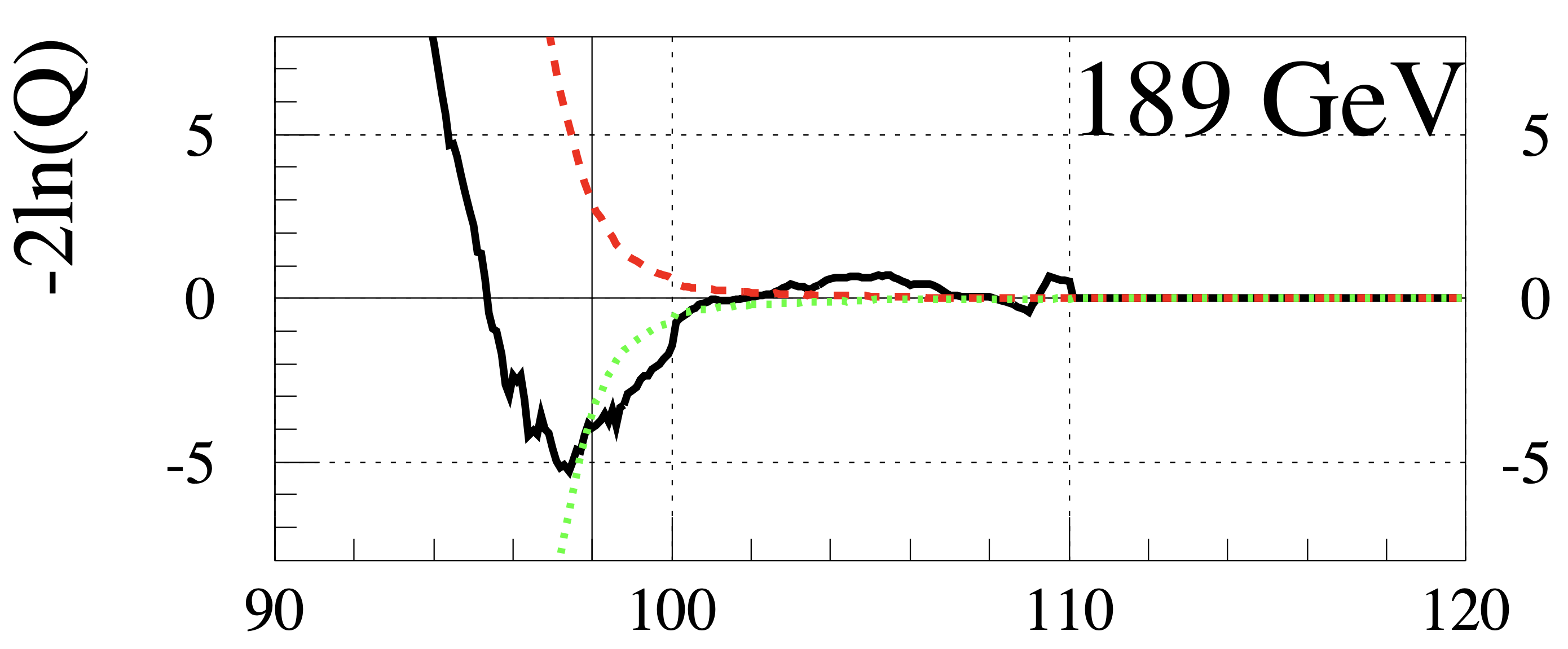}
    \caption{\small (From Ref.~\cite{alephcollaboration2001search}) The observed (negative) log-likelihood ratio at $\sqrt{s}=189$\,GeV as a function of the test Higgs boson mass (full curve). The dashed/dotted curves show the expected average behaviour as a function of the true Higgs boson mass for background and signal+background.}
    \label{fig:ExpLik189}
\end{figure}

On the sole basis of these rates (thus ignoring the details of reconstructed mass distribution), the observed excess appears to be compatible with a 98\,GeV SM Higgs boson. This statement is confirmed by the observed behaviour of the (negative) log-likelihood ratio as a function of the {\it test} Higgs boson mass, displayed in Fig.~\ref{fig:ExpLik189}~\cite{alephcollaboration2001search}. This test-statistics compares the compatibility of the data with the background and the signal-plus-background hypotheses, on the basis of the number of events and the {\it reconstructed} mass distribution, but also the jet flavour tagging, to mention only the most discriminant variables. This log-likelihood ratio exhibits a minimum for a {\it test} mass around 98\,GeV, in agreement with the value expected from the signal-plus-background hypothesis at this {\it true} mass\footnote{At this point, the reader's attention is called to reflect upon the difference between the {\it reconstructed}, the {\it test} and the {\it true} Higgs boson masses. A clear mastering of these totally different concepts, albeit mixed up in Figs.~\ref{fig:ObsMass189} and~\ref{fig:ExpLik189}, is required to avoid claims such as those made in Refs.~\cite{Biekotter:2022jyr,Biekotter:2023jld} and the following avatars.}.

On the other hand, a 95\,GeV scalar with 0.117 signal strength, in line with what is requested from Refs.~\cite{Aguilar-Saavedra:2023tql,Ahriche:2023wkj,Ahriche:2023hho,Arcadi:2023smv,Arhrib:2024wjj,Ashanujjaman:2023etj,Ayazi:2024fmn,Azevedo:2023zkg,Belyaev:2023xnv,Belyaev:2024lah,Benbrik:2024ptw,Biekotter:2022jyr,Biekotter:2023jld,Biekotter:2023oen,Borah:2023hqw,Cao:2023gkc,Cao:2024axg,Chen:2023bqr,Dev:2023kzu,Diaz:2024yfu,Dutta:2023cig,Ellwanger:2023zjc,Ellwanger:2024txc,Ellwanger:2024vvs,Escribano:2023hxj,Ge:2024rdr,Kalinowski:2024oyy,Kalinowski:2024uxe,Li:2023kbf,Lian:2024smg,Liu:2024cbr,Wang:2024bkg,Yang:2024fol,Mosala:2024mcy,Gao:2024qag} to support the LHC findings at 95\,GeV, would be able to account for almost half of the observed LEP excess,\footnote{A signal strength of 0.28 would be needed for a 95\,GeV scalar to account for the entirety of the excess observed.} and cannot be excluded. The atypical observed mass distribution of the excess, however, almost uniformly distributed from 87 to 99\,GeV, led the LEP experiments and the LEP Higgs working group to comment as follows in Ref.~\cite{Bock:411797}: ``{\it The observation appears to deviate from the SM background expectation over a large range of test masses, {\rm [in a way that]} is not typical of a Higgs boson signal but indicates, rather, a slight overall excess of the data with respect to the SM background prediction (also apparent in Fig.~\ref{fig:ObsMass189}) which could either be a statistical fluctuation or a systematic underestimate of the {\rm [ZZ]} background.}''. 

To firmly decide whether the LEP excess either arises from a small background fluctuation or a 98\,GeV SM Higgs boson signal, or is a hint of a 95\,GeV signal with strength 0.117 requires significantly more data. It so happens, as shown in Table~\ref{tab:LEPLumis}, that three times more data were collected by the LEP experiments in 1999 and 2000 at higher centre-of-mass energies (from 192 to 210\,GeV), with the additional bonus of a larger ZH cross section (further away from the ZH production threshold). Should the excess observed at $\sqrt{s} = 189$\,GeV be due to a genuine 95 (98)\,GeV signal, about 5 (13) times more signal events would have been produced in the 1999/2000 data, while the dominant ZZ background increases only by a factor of four. The observed 2.3 standard deviations at 189\,GeV should therefore naively become an unambiguous discovery with $5.8\sigma$ ($15\sigma$) significance in the 1999/2000 data -- which has not happened. These data are examined in a more quantitative manner in the next section.

 \section{What do the LEP data at \texorpdfstring{$\sqrt{s} \ge 192$\,GeV}{192} say?}
 \label{sec:192-210}

\begin{figure}[htbp]
    \centering
    \includegraphics[width=0.60\textwidth]{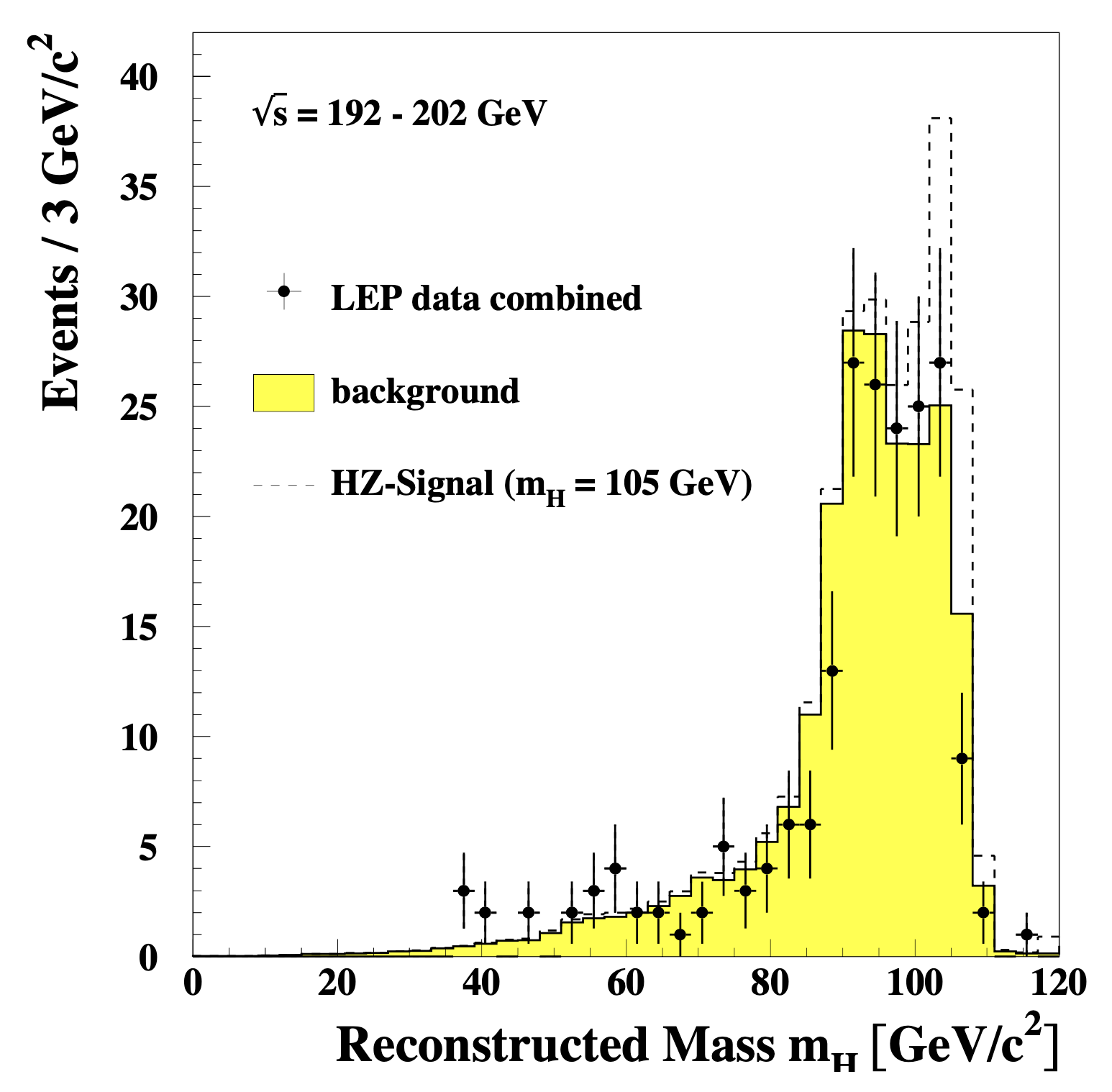}
    \caption{\small (From Ref.~\cite{Carena:436918}) Distribution of the reconstructed mass in SM Higgs boson searches conducted at $\sqrt{s}$ between 192 and 202\,GeV. The figure displays the data (dots with error bars), the predicted SM background (shaded histogram) and the prediction for a Higgs boson of 105\,GeV mass (dashed histogram). The number of data events in this figure is 201 while 220 are expected from SM background processes. A signal at 105\,GeV would contribute 41 additional events.}
    \label{fig:ObsMass192-202}
\end{figure}

A first update of the SM Higgs boson search at LEP was prepared for the Winter 2000 conferences in Ref.~\cite{Carena:436918}, with the data taken in 1999 at centre-of-mass energies between 192 and 202\,GeV. The {\it reconstructed} mass distribution obtained with these data is displayed in Fig.~\ref{fig:ObsMass192-202}, together with the expectation from a 105\,GeV SM Higgs boson. The data agree with the SM background over the whole reconstructed mass range. 

\noindent The following observations can be made.
\begin{enumerate}
\item As indicated in Ref.~\cite{Carena:436918}, a signal at 105\,GeV would contribute 41 events in this distribution, out of 77 events expected to be produced with $\rm H\to b\bar b$. Of these 41 events, 31 (24) are expected to be reconstructed with a mass within a $\pm 6\,(\pm 3) $\,GeV window around 105\,GeV, corresponding to a selection efficiency of 40 (31)\% in this mass range. These efficiencies are not expected to vary significantly with the true Higgs boson mass. 
\item With such selection efficiency, an excess of 80 events would be expected from a 98\,GeV SM Higgs boson in a 12\,GeV reconstructed mass interval around 98\,GeV. In this interval, 102 events are observed, in agreement with 99 events expected from the SM background. The probability that $99+80 = 179$ events expected fluctuates down to 102 observed events or less is about $3\times 10^{-10}$. A 98\,GeV SM Higgs boson is therefore excluded beyond any doubts (with 99.99999997\% confidence) from these 1999 data alone. This interpretation of the excess observed in the 1998 data is thus dismissed in the following. 
\item Similarly, an excess of 12 (9) events would be expected from a 95\,GeV scalar with a signal strength of 0.117 in a 12 (6)\,GeV reconstructed mass interval around 95\,GeV. In this interval, 102 (50) events are observed in agreement with 102 (51) events expected from the SM background. The existence of such a 95\,GeV signal is therefore not upheld by the 1999 data. These data alone, however, do not suffice to reach a 95\% confidence level exclusion. The probability that $51+9=60$ events expected fluctuates down to 50 events observed or less is still 9.3\%. The 95\,GeV interpretation of the excess observed in the 1998 data is disfavoured with more than 90\% C.L., but the 2000 data are needed to go further. 
\end{enumerate}
 
\begin{figure}[htbp]
    \centering
    \includegraphics[width=0.90\textwidth]{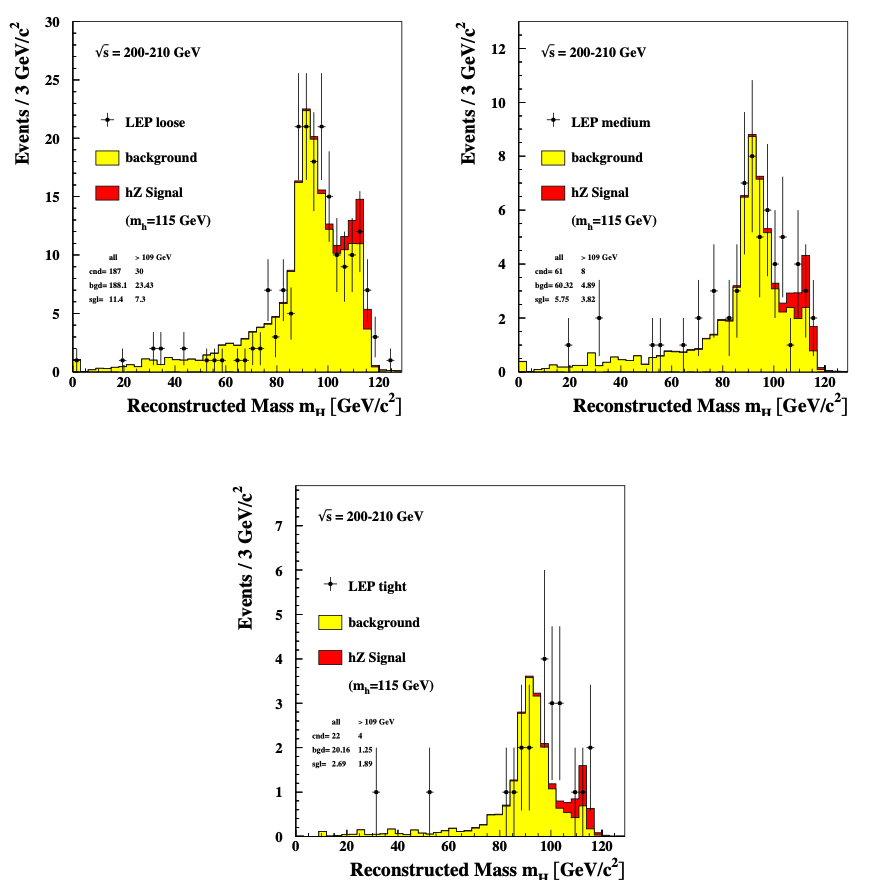}
    \caption{\small Distributions of the reconstructed Higgs boson mass obtained from three non-biasing selections with increasing signal purities, from the LEP data collected in the year 2000. The histograms show the Monte Carlo predictions, lightly shaded for the SM background, heavily shaded for an assumed SM Higgs boson of mass 115\,GeV, together with the data (dots with error bars). The numbers of events observed and expected are given in the three plots.}
    \label{fig:ObsMass202-209}
\end{figure}

Shortly after the end of the LEP data taking in November 2000, the LEP Higgs working group updated again their findings~\cite{alephcollaboration2001search}, with the data taken in the year 2000. The three reconstructed mass distributions shown in this paper with three different expected signal purities -- called "Loose", "Medium" and "Tight" -- are reproduced in Fig.~\ref{fig:ObsMass202-209}, together with the expectation from the SM backgrounds and from a 115\,GeV SM Higgs boson. A small excess of events was observed at this mass in the last few months of the year 2000 when LEP was producing collisions at the highest energies (above 206\,GeV). The LHC search for the SM Higgs boson search showed in 2011 that this excess was also due to a fluctuation of the SM background.   
The numbers of events observed in a 12\,GeV reconstructed mass interval around 95\,GeV are listed in Table~\ref{tab:N200-210} for each of the selections, together with the number of events expected from SM backgrounds. On the basis on the numbers of signal expected for a Higgs boson mass of 115\,GeV, the selection efficiencies for a 95\,GeV signal\footnote{Unlike in the final publication of the LEP Working Group~\cite{LEPWorkingGroupforHiggsbosonsearches:2003ing}, non-biasing selection algorithms are used in Ref.~\cite{alephcollaboration2001search} for these three mass distributions.} in this mass interval are estimated to be of the order of 40\%, 20\% and 10\%, respectively, yielding the number of signal events expected for a signal strength of 0.117 as displayed in the last row of Table~\ref{tab:N200-210}. 
\begin{table}[htbp]
    \renewcommand{\arraystretch}{1.3}
    \centering
    \caption{\small For each of the three (Loose, Medium, Tight) selections applied to the LEP data collected in 2000 in a 12\,GeV mass interval around 95\,GeV: number of events observed (first row); number of events expected from the SM backgrounds (middle row); number of signal events expected for a mass of 95\,GeV and a signal strength of 0.117 (last row). 
    }
    \begin{tabular}{|c|c|c|c|}
    \hline
    Selection  & Loose & Medium & Tight \\ \hline
    Number of events observed & 75 &  23 & 9  \\ \hline
    Number of background events expected & 71 & 24 & 10   \\ \hline
    Number of signal events expected & 13 &  6.5 & 3.2    \\ \hline
    \end{tabular}
    \label{tab:N200-210}
\end{table}

The data are well described by the Standard Model predictions in this mass interval. The numbers of events observed in the three selections certainly do not suggest the presence of a 95\,GeV scalar and a signal strength of 0.117. Because the ``Loose'' selection of the 2000 data has an efficiency similar to that of the selection used in Fig.~\ref{fig:ObsMass192-202} for the 1999 data, it is chosen for a combination of all data at $\sqrt{s} \ge 192$\,GeV. In the 1999/2000 combined data, an excess of 25 (19) events would be expected from a 95\,GeV scalar with a signal strength of 0.117 in a 12 (6)\,GeV reconstructed mass interval around 95\,GeV. In this interval, 177 (89) events are observed in agreement with the 173 (86) events expected from the SM background. The probability that $86+19= 105$ events expected in a 6\,GeV window around 95\,GeV fluctuates down to 89 events observed or less is reduced to 6.2\%. The interpretation of the LEP excess at $\sqrt{s}=189$\,GeV as originating from a 95\,GeV scalar particle with a signal strength of 0.117 is therefore excluded at the 94\% C.L. by the subsequent data. 

Now, as mentioned in Section~\ref{sec:189}, the LEP data at $\sqrt{s}=189$\,GeV would actually require a larger signal strength of $\sim 0.28$ for a 95\,GeV scalar particle to account for the totality of the excess observed with reconstructed masses above 90\,GeV. It is only when the 1998 data are combined with the 1999/2000 data -- where there is no evidence for such a particle -- that the (internally inconsistent) 95\,GeV signal strength artificially shrinks to 0.117. Such a signal strength of 0.28 is excluded at the 99.994\% C.L. by the 1999/2000 data.  The same exercise can be repeated for all masses between 95 and 100\,GeV, a range sometimes referred to in new physics interpretations. The signal strength needed to entirely account for the excess observed at $\sqrt{s}=189$\,GeV, is shown in Table~\ref{tab:excess189} as a function of the hypothetical scalar particle mass, together with the number of signal events expected in a 12\,GeV mass interval and the level of exclusion of this hypothesis with the subsequent data at $\sqrt{s} \ge 192$\,GeV. 

\begin{table}[htbp]
    \renewcommand{\arraystretch}{1.3}
    \centering
    \caption{\small For scalar masses between 95 and 100\,GeV: signal strength needed to entirely account for the excess observed at $\sqrt{s} = 189$\,GeV with reconstructed masses above 90\,GeV; number of signal events expected in a 12\,GeV mass interval around the hypothesised scalar mass; probability to observe as many events or less in this mass interval with the data sample at $\sqrt{s} \ge 192$\,GeV under the signal hypothesis. 
    }
    \begin{tabular}{|c|c|c|c|c|c|c|}
    \hline
    Mass (GeV)  & 95 & 96 & 97 & 98 & 99 & 100 \\ \hline
    Signal strength & 0.28 &  0.37 & 0.53 & 0.85 & 1.33 & 1.88  \\ \hline
    Signal events expected & 61 & 74 & 99 & 150 & 221 & 290 \\ \hline
    Signal probability & $6.0\,10^{-5}$ & $1.7\,10^{-6}$ & $5.0\,10^{-10}$ & $3.5\,10^{-17}$ & $1.6\,10^{-32}$ & $5.6\,10^{-50}$\\ \hline
    \end{tabular}
    \label{tab:excess189}
\end{table}

As a consequence, the most likely and only remaining explanation of the LEP b\=b excess with reconstructed mass above 90\,GeV, {\it observed solely at $\sqrt{s} =$ 189\,GeV}, is either a small upward statistical fluctuation or an underestimate of the ZZ background at this centre-of-mass energy. A similar, though less quantitative, statement had already been made by the LEP experiments and the LEP Higgs working group in Ref.~\cite{alephcollaboration2001search}: ``{\it In the 189\,GeV data, an excess had been observed [...] compatible with the dominant $e^+e^- \to ZZ$ background. [...] There is no evidence for a systematic effect at threshold in the data collected above 189\,GeV.}''.

\section{Conclusion}
\label{sec:conclusion}

In this note, it has been recalled that a 2.3$\sigma$ excess ($\sim 12$ events) over the SM backgrounds is observed in the Higgs boson search at LEP within the 683\,pb$^{-1}$ of data collected in 1998 at $\sqrt{s} = 189$\,GeV. While this excess is compatible in strength with the production of a Standard Model Higgs boson of mass 98\,GeV, three popular interpretations (either a small upward fluctuation of the SM backgrounds, or a 98\,GeV SM Higgs boson, or a 95\,GeV scalar with reduced signal strength) were still open at the end of 1998. For the latter interpretation, a signal strength of 0.28 is needed to account for the observed excess, reduced to 0.117 when the data collected in 1999 and 2000, where no such deviation is observed, are included. A careful scrutiny of the reconstructed mass distributions made with the data sample of 1776\,pb$^{-1}$ collected at $\sqrt{s}$ between 192\,GeV and 209\,GeV in 1999 and 2000 has indeed not revealed any excess of events compatible with the last two interpretations of the small excess seen in 1998. The 1999 data alone exclude the 98\,GeV scalar interpretation with a level of confidence of 99.99999997\%. The combination of the 1999 and 2000 data excludes the 95\,GeV scalar interpretation with a signal strength of 0.117 (0.28) at the 94\% (99.994\%) confidence level. 

\begin{figure}[htbp]
    \centering
    \includegraphics[width=0.43\textwidth]{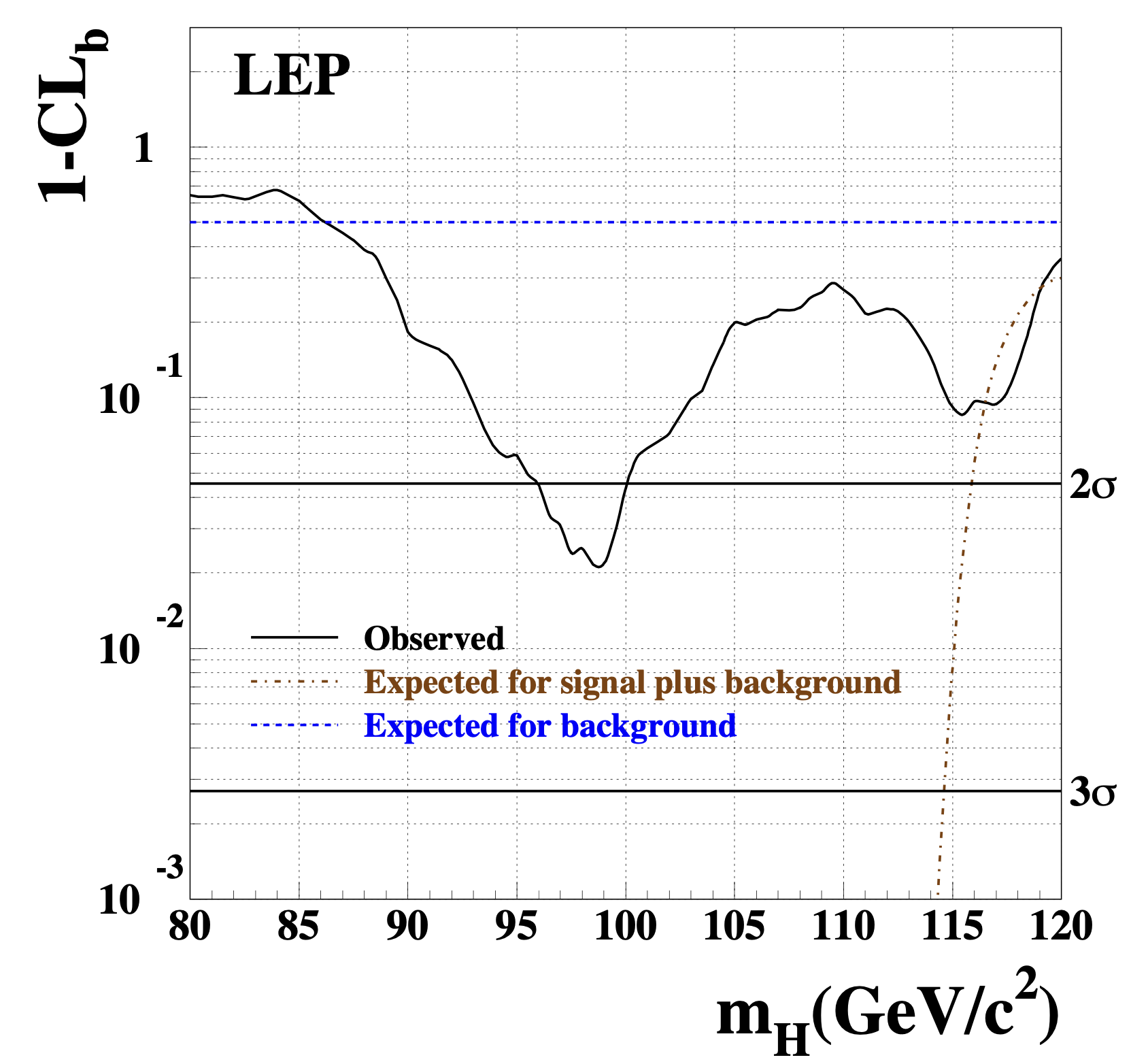}
    \includegraphics[width=0.55\textwidth]{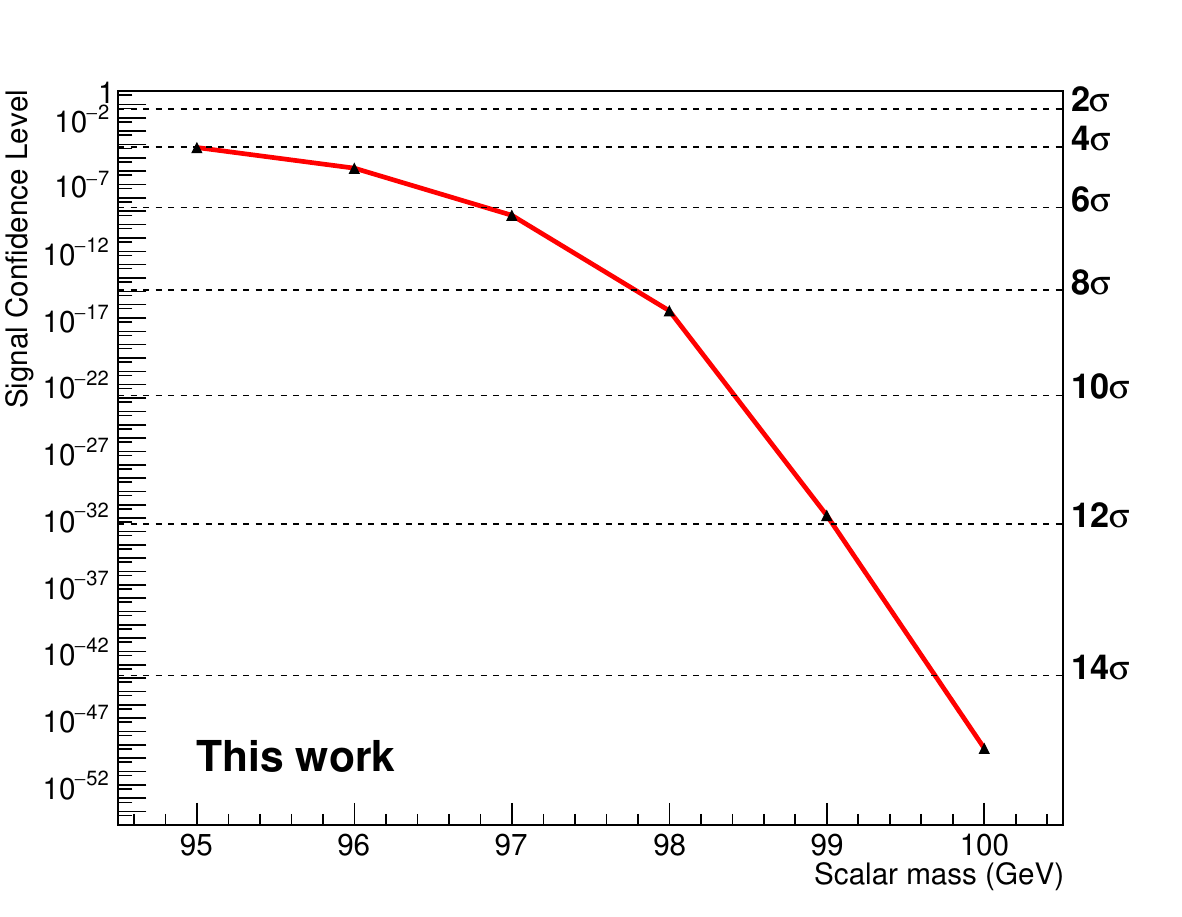}
    \caption{\small Left panel, taken from Ref.~\cite{LEPWorkingGroupforHiggsbosonsearches:2003ing}: The LEP-combined background confidence level $1-{\rm CL_b}$ as a function of the test mass $m_{\rm H}$, with all data at centre-of-mass energies from 189 to 209\,GeV. Full curve: observation; dashed curve: expected background confidence level; dash-dotted line: the position of the minimum of the median expectation of $1-{\rm CL_b}$ for the signal plus background hypothesis, when the signal mass indicated on the abscissa is tested. The horizontal solid lines indicate the levels for 2$\sigma$ and 3$\sigma$ deviations from the background hypothesis. Right panel, this work: The signal (plus background) hypothesis confidence level as a function of the test mass from 95 to 100\,GeV, with data at centre-of-mass energies above 192\,GeV. The signal strength is tuned as a function of the test mass to account for the excess observed at $\sqrt{s} = 189$\,GeV. The horizontal solid lines indicate the levels for 2$\sigma$ to 14$\sigma$ deviations from the signal-plus-background hypothesis.
    \label{fig:ClbCombined}}
\end{figure}

In the defence of the authors of Refs.~\cite{Aguilar-Saavedra:2023tql,Ahriche:2023wkj,Ahriche:2023hho,Arcadi:2023smv,Arhrib:2024wjj,Ashanujjaman:2023etj,Ayazi:2024fmn,Azevedo:2023zkg,Belyaev:2023xnv,Belyaev:2024lah,Benbrik:2024ptw,Biekotter:2022jyr,Biekotter:2023jld,Biekotter:2023oen,Borah:2023hqw,Cao:2023gkc,Cao:2024axg,Chen:2023bqr,Dev:2023kzu,Diaz:2024yfu,Dutta:2023cig,Ellwanger:2023zjc,Ellwanger:2024txc,Ellwanger:2024vvs,Escribano:2023hxj,Ge:2024rdr,Kalinowski:2024oyy,Kalinowski:2024uxe,Li:2023kbf,Lian:2024smg,Liu:2024cbr,Wang:2024bkg,Yang:2024fol,Mosala:2024mcy,Gao:2024qag}, it remains true that the confidence level of the background-only hypothesis $1-{\rm CL_{\rm b}}$ obtained with a combination of all data from 189 to 209\,GeV presents a minimum at 98\,GeV, still corresponding to a $2.3\sigma$ deviation (with no increase whatsoever with respect to the sole 189\,GeV data). The distribution of this confidence level as a function of the {\it test} mass is shown in the left panel of Fig.~\ref{fig:ClbCombined}, reproduced from the final LEP Higgs working group publication in 2003~\cite{LEPWorkingGroupforHiggsbosonsearches:2003ing}. 
Two independent small excesses conspire to produce this small deviation. First, an excess of events is observed above 102\,GeV (7 events observed vs 2.4 expected) in the Tight selection of Fig.~\ref{fig:ObsMass202-209} from the data collected in the last few months of LEP, which leads to the famous secondary minimum at 115\,GeV in Fig.~\ref{fig:ClbCombined}. This excess biases the background confidence level to smallish values (at the level of $\sim 1\sigma$) all the way down to {\it test masses} of 90\,GeV because of non-Gaussian tails in the {\it reconstructed mass} distributions expected for these {\it true masses}. This first excess has been demonstrated by the LHC to be due to a statistical fluctuation of the background rather than to the production of a 115\,GeV Higgs boson. Second, the excess observed at $\sqrt{s} = 189$\,GeV sculpts a minimum around a test mass of 98\,GeV, increasing the overall deviation from $\sim 1\sigma$ to over $2\sigma$ in the {\it test mass} interval between 95 and 100\,GeV.\footnote{Back in June 2000, that is before the appearance of this excess at reconstructed masses above 102\,GeV, the combined $1-CL_{\rm b}$ was contained between 30\% and 70\% for any test mass above 102\,GeV, and amounted to 20\% at 100\,GeV~\cite{LEPCollaborations:2000txm}. The high-mass excess made it drop to 5\% at 100\,GeV, as shown in Fig.~\ref{fig:ClbCombined}, i.e., from less than $1\sigma$ to about $2\sigma$.} 
Should this excess observed at 189\,GeV be due to the production of a scalar particle with mass between 95 and 100\,GeV, it would have been backed up by larger excesses in the data collected at higher centre-of-mass energies. Such excesses have not been witnessed. The right panel of Fig.~\ref{fig:ClbCombined} shows the confidence level of this signal (plus background) hypothesis in this mass range with a combination of all data from 192 to 209\,GeV (also in Table~\ref{tab:excess189}). The signal-plus-background hypothesis is at least 1000 times less likely than a background-only fluctuation at 189\,GeV.

Sadly, two background fluctuations in very different mass ranges do not make a new physics signal. 
It is therefore high time to stop using these fluctuations in support of any signal interpretation of the $3\sigma$ excess observed around 95\,GeV by CMS in their diphoton mass distribution. Altogether, the 1999-2000 LEP data strongly disfavour the production of a new 95\,GeV Higgs boson with a signal strength of 0.117, as well as any other new physics interpretation in the 95--100\,GeV mass range of the 2.3$\sigma$ excess observed in the 1998 data .

\section*{Disclaimer}

{The numbers presented in this note have been obtained solely from the information publicly available in Refs.~\cite{Bock:411797, Carena:436918, alephcollaboration2001search,LEPWorkingGroupforHiggsbosonsearches:2003ing, Mele:456346, LEPCollaborations:2000txm}, and with the help of the HZHA generator~\cite{hzha} for the Higgs production cross sections and branching fractions at $m_{\rm H} = 91, 95, 98, 105$ and $115$\,GeV. In particular, the numbers of events observed and expected in specific mass distribution intervals have been directly read off the histograms of Figs.~\ref{fig:ObsMass189},~\ref{fig:ObsMass192-202} and~\ref{fig:ObsMass202-209}. The expected numbers were rounded to the nearest integer. Other event-by-event information, used by the LEP Higgs working group 20 years ago, was not available to the author. Should this event-by-event information still exist, exclusion confidence levels could be further optimized with a refined analysis carried out by former members of the LEP Higgs working group. The conclusions of the reported (sub-optimal, but statistically sound) analysis, which do not challenge the published LEP limits on Higgs boson production but demonstrate the incompatibility of the data at $\sqrt{s} \geq 192$\,GeV with any signal interpretation of the small excess observed at 189 GeV with large confidence levels, are probably already clear enough.}      

\acknowledgments
I can't thank Daniel Treille enough for giving me the inspiration and the motivation to start the archaeological work presented in this note. It would be unfair not to gratefully mention the presentation of Thomas Biek\"otter, Sven Heinemeyer and Georg Weiglein in the CERN Collider Cross Talk meeting~\cite{CCT} and their subsequent contribution to the CERN/EP Newsletter~\cite{Newsletter}, which made me bite the bullet. I have also no alternative but to warmly thank Christophe Grojean and Jean-Baptiste de Vivie for a careful reading of the manuscript and for many subtle suggestions. Finally, my best thoughts go to Marumi Kado and Jean-Baptiste de Vivie, who had written 25 years ago an internal ALEPH note entitled ``Two $\tilde{\rm b}$ or not two $\tilde{\rm b}$?''~\cite{Kado:647451}, a title that I shamelessly plagiarised here.


\bibliographystyle{JHEP}
\bibliography{biblio}

\end{document}